\documentstyle[11pt,aas2pp4,psfig]{article}

\begin{document}

\lefthead{Psaltis et al.}
\righthead{Kilohertz Quasi-Periodic Oscillations}

\title{The Beat-Frequency Interpretation of Kilohertz 
QPOs\\ in Neutron Star Low-Mass X-ray Binaries}

\author{Dimitrios Psaltis\altaffilmark{1},
Mariano M\'endez\altaffilmark{2,3},
Rudy Wijnands\altaffilmark{2},
Jeroen Homan\altaffilmark{2},
Peter G.\ Jonker\altaffilmark{2},\\
Michiel van der Klis\altaffilmark{2},
Frederick K.\ Lamb\altaffilmark{4},
Erik Kuulkers\altaffilmark{5},
Jan van Paradijs\altaffilmark{2,6},\\
and Walter H.\,G.\ Lewin\altaffilmark{7}}

\begin{center}
To appear in {\em The Astrophysical Journal Letters\/}.
\end{center}

\altaffiltext{1}{Harvard-Smithsonian Center for
Astrophysics, 60 Garden St., Cambridge, MA 02138,
U.S.A.; dpsaltis@cfa.harvard.edu}

\altaffiltext{2}{Astronomical Institute ``Anton
Pannekoek'', University of Amsterdam and Center for High
Energy Astrophysics, Kruislaan 403, NL-1098 SJ Amsterdam,
The Netherlands; mariano, rudy, homan, peterj, michiel, 
jvp@astro.uva.nl}

\altaffiltext{3}{Facultad de Ciencias Astron\'omicas 
y Geofisicas, Universidad Nacional de la Plata, Pasco 
del Bosque S/N, 1090 La Plata, Argentina}

\altaffiltext{4}{Departments of Physics and Astronomy,
University of Illinois at Urbana-Champaign, 1110 W. Green
St., Urbana, IL 61801, U.S.A.; f-lamb@uiuc.edu}

\altaffiltext{5}{Astrophysics, University of Oxford,
Nuclear and Astrophysics Laboratory, Keble Road,
Oxford, OX1 3RH, United Kingdom;
e.kuulkers1@physics.oxford.ac.uk}

\altaffiltext{6}{Department of Physics, University of
Alabama at Huntsville, Huntsville, AL 35899, U.S.A.}

\altaffiltext{7}{Department of Physics and Center for
Space Research, Massachusetts Institute of Technology,
Cambridge, MA 02139, U.S.A.; lewin@space.mit.edu}
 
\vspace{1.0cm}

\begin{abstract}
 Pairs of quasi-periodic oscillations (QPOs) at kilohertz frequencies are
a common phenomenon in several neutron-star low-mass X-ray binaries. The
frequency separation of the QPO peaks in the pair appears to be constant
in many sources and directly related to the neutron star spin frequency.
However, in Sco~X-1 and possibly in 4U~1608$-$52, the frequency separation
of the QPOs decreases with increasing inferred mass accretion rate. We
show that the currently available {\em Rossi X-ray Timing Explorer\/} data
are consistent with the hypothesis that the frequency separations in all
sources vary by amounts similar to the variation in Sco~X-1. We discuss
the implications for models of the kilohertz QPOs. 
 \end{abstract}





\section{INTRODUCTION}

Quasi-periodic X-ray brightness oscillations at
kilohertz frequencies (hereafter kHz QPOs) have
recently been discovered in many neutron-star low-mass
X-ray binaries (LMXBs) with the {\em Rossi X-ray
Timing Explorer\/} (RXTE; see, e.g., van der Klis et
al.\markcite{vdketal96} 1996; Strohmayer et
al.\markcite{Setal96} 1996). These are strong, often
relatively coherent ($\nu/\delta \nu$ up to $\sim
200$) oscillations that occur commonly in pairs (see
van der Klis\markcite{vdk98} 1998 for a recent
review). 

The frequencies of the kHz QPOs are comparable to the
dynamical timescale near the neutron star surface and
depend on the mass accretion rate as inferred from the
observed countrates and the spectral properties of the
sources (van der Klis et al.\ 1996; Strohmayer et al.\
1996\markcite{Setal96}; Ford et
al.\markcite{Fetal97a}\markcite{Fetal97b} 1997a,
1997b;  van der Klis et al.\markcite{vdketal97} 1997).
The peak separation between the lower-frequency
(hereafter the lower kHz QPO) and the upper-frequency
kHz QPO (hereafter the upper kHz QPO) in a given
source is generally consistent with a constant value,
independent of the mass accretion rate (Strohmayer et
al.\ 1996; Ford et al.\ 1997a, 1997b; Wijnands et
al.\markcite{Wetal97b}\markcite{Wetal98a}\markcite{Wetal98b}
1997b, 1998a, 1998b). In 4U~1728$-$34 and in
4U~1702$-$43, this peak separation is closely equal to
the frequency of the nearly coherent oscillations
observed during type~I X-ray bursts that are thought
to be produced at the spin frequencies of the neutron
stars (Strohmayer et al.\ 1996;  Strohmayer, Zhang, \&
Swank\markcite{SZS97}\markcite{SZS98} 1997, 1998); in
4U~1636$-$536 and in KS~1731$-$26, the peak separation
is closely equal to half the frequency of the nearly
coherent oscillations observed during type~I X-ray
bursts (Smith, Morgan, \& Bradt\markcite{SMB97} 1997;
Wijnands \& van der Klis\markcite{WK97} 1997;
Strohmayer et al.\ 1998). 
 
The above observations offer strong evidence in favor
of beat-frequency models, in which the frequency of
the lower kHz QPO is the beat frequency between the
upper kHz QPO and the neutron star spin (Strohmayer et
al.\ 1996; Miller, Lamb, \& Psaltis\markcite{MLP98}
1998). In Sco~X-1, however, which is a luminous LMXB,
the peak separation of the kHz QPOs is {\em not\/}
constant, but decreases with increasing inferred mass
accretion rate (van der Klis et al.\ 1997). In
4U~1608$-$52, which is a less luminous LMXB, there is
also evidence for a peak separation that is not
constant (M\'endez et al.\markcite{Metal98} 1998). 

In this paper, we use previously published {\em
RXTE\/} data on several low-mass X-ray binaries to
critically discuss the evidence in favor of a constant
frequency separation between the kHz QPO peaks
required in any simple beat frequency interpretation.
We find that the current data on all sources except
Sco~X-1 are insufficient, when used individually, to
distinguish between a constant peak separation and a
peak separation that varies by amounts similar to
those seen in Sco~X-1. When we use the combined
dataset of all sources, we find a remarkable
correlation between the frequencies of the lower and
upper kHz QPOs, which suggests that the peak
separation may be varying in all sources. 

\section{IS THE PEAK SEPARATION CONSTANT?}

The current data on kHz QPOs include observations of
eleven neutron-star LMXBs in which pairs of kHz QPOs
have beed detected in their persistent emission. In
three of these systems, nearly coherent oscillations
have been detected during type~I X-ray bursts at
frequencies consistent with being equal to the peak
separation between the kHz QPOs or their first
overtones. In 4U~0614+09, a third QPO has been
detected with marginal significance (reported to be
$\sim 2.7\sigma$; Ford et al.\ 1997a) at a frequency
consistent with being equal to the peak separation
between the kHz QPOs. Table~1 summarizes these
observations. 

\begin{figure*}[t]
\centerline{
\psfig{file=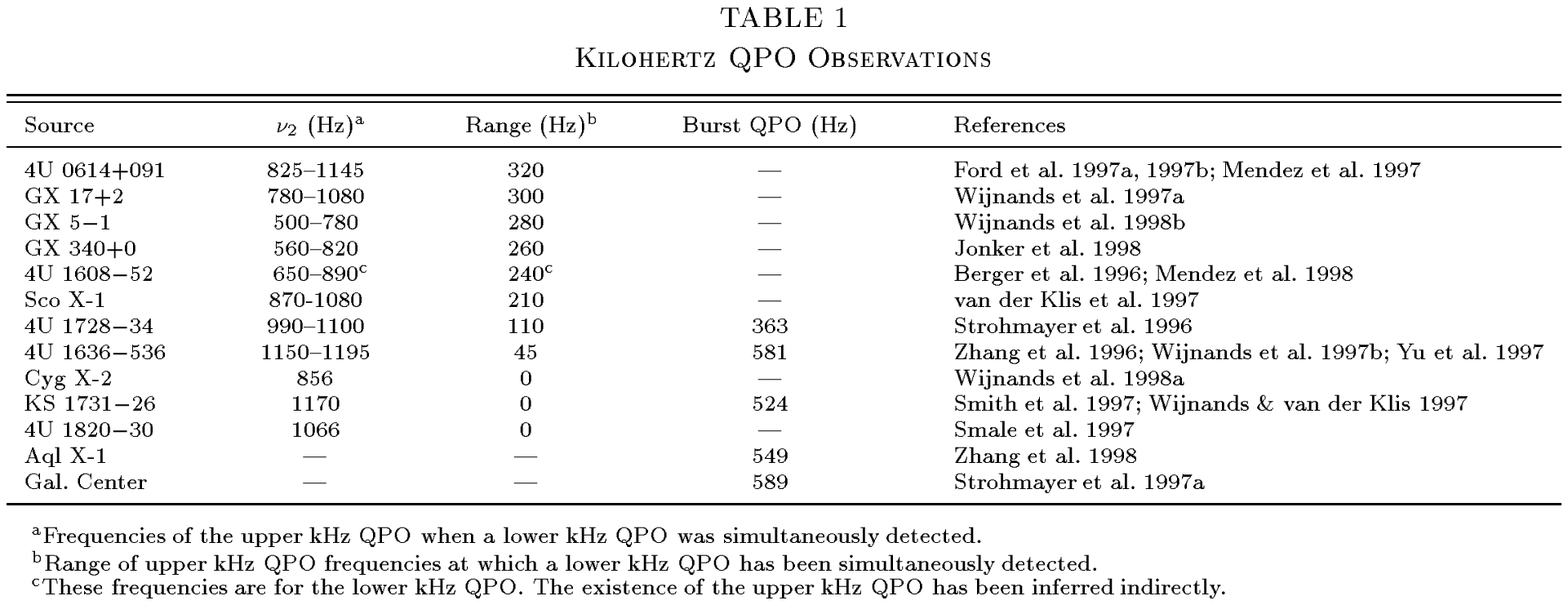,angle=0,height=5.5cm,width=450pt}}
\end{figure*}

Testing the hypothesis of a constant peak separation
between the kHz QPOs requires for each individual
source the detection of both QPOs over a wide range of
frequencies. In the three sources in which both the
pair of kHz QPOs in the persistent emission and nearly
coherent oscillations during type~I X-ray bursts have
been detected, the range of reported frequencies of
simultaneously detected kHz QPO pairs in the
persistent emission is {\em very\/} narrow (see
Table~1). In all sources, besides Sco~X-1, in which
the range of detected frequencies of the pairs of kHz
QPOs is wide ($\gtrsim 200$~Hz in GX~17$+$2,
GX~340$+0$, GX~5$-$1, and 4U~0614$-$09), the
fractional errors in the measurement of the centroid
frequencies of the QPOs are substantial. Figure~1
compares the distribution of fractional ($1\sigma$)
errors in the determination of the peak separation
between the kHz QPOs in Sco~X-1 with the corresponding
distribution for GX~17$+$2, GX~340$+0$, GX~5$-$1, and
4U~0614$-$09. In all sources besides Sco~X-1, the
average 1$\sigma$ error is $\sim$10--50~Hz, comparable
to the change in the peak separation observed in
Sco~X-1. 

\begin{figure}[t]
\centerline{
\psfig{file=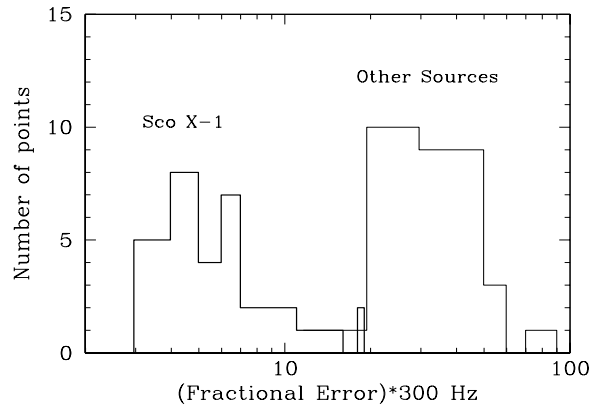,angle=-90,height=5.5truecm,width=8.0truecm}}
 \figcaption[] {Distribution of fractional errors in
the determination of the peak separation between the
kHz QPOs in Sco~X-1 and in GX~17$+$2, GX~340$+0$,   
GX~5$-$1, and 4U~0614$-$09.}
\end{figure}

Figure~2a shows the results of testing the simple
beat-frequency interpretation of the pairs of kHz QPOs
in several sources by means of a $\chi^2$ test.  We
performed this test for the five sources in our sample
for which the range of frequencies of simultaneously
detected upper and lower kHz QPOs is wide (i.e., at
least 200~Hz) using for each source all the
measurements reported in the references listed in
Table~I; the uncertainties in the measurements were
estimated in the same way for all sources and hence
our observational sample is uniform. The more limited
data on 4U~1728$-$34 and 4U~1636$-$536 are consistent
with the result presented in Figure~2. Figure~2a shows
the resulting $\chi^2$ values for these sources,
confirming the known result that the Sco~X-1 data are
inconsistent with a constant peak separation, whereas
the data on the remaining sources are consistent with
it (see also van der Klis et al.\ 1997; Wijnands et
al.\ 1997b, 1998a, 1998b;  Ford et al.\ 1997a, 1997b).
In GX~5$-$1, the minimum $\chi^2$ value is $\sim 2$
mostly because we included for this source frequencies
of kHz QPOs that were marginally detected (at
$<3\sigma$; see Wijnands et al.\ 1998b for a
discussion). The peak separation in all sources is
$\sim 300$~Hz. 

\begin{figure}[t]
\centerline{
\psfig{file=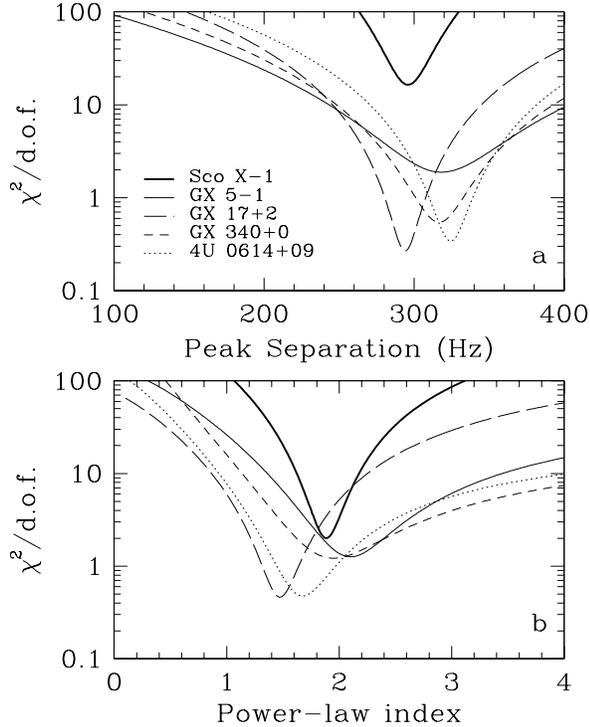,angle=0,height=10truecm,width=8.0truecm}}
 \figcaption[] {Results of testing (a) the simple
beat-frequency interpretation of the pairs of kHz QPOs
and (b) the hypothesis of a power-law correlation
between the frequencies of kHz QPOs in five LMXBs by
means of a $\chi^2$ test.}
\end{figure}

The peak separation between the kHz QPOs in Sco~X-1
decreases with increasing inferred mass accretion rate
(van der Klis et al.\ 1997). Figure~3a shows the
frequency $\nu_1$ of the lower kHz QPO in Sco~X-1
plotted against the frequency $\nu_2$ of the upper kHz
QPO. The data points can be adequately described by
the power-law relation
 \begin{equation}
 \nu_1= (724\pm 3) 
  \left(\frac{\nu_2}{1000~\mbox{Hz}}\right)^{1.9\pm 
      0.1}~\mbox{Hz}\;,
 \label{scopl}
 \end{equation}
 which is shown in the figure as a dashed line. The
$\chi^2$ value that corresponds to
relation~(\ref{scopl}) is 2.0 (see Figure~2b). This is
a relatively high value because, at high ($\sim
1100$~Hz) frequencies of the upper kHz QPO, the QPO
peaks are weak and therefore the measurements are
dominated by systematic uncertainties; when we use
only the data points with upper kHz QPO frequency $\le
1050$~Hz, the $\chi^2$ value for
relation~(\ref{scopl}) becomes $\simeq 1.5$. It is
important to note that the power-law description was
arbitrarily chosen here to describe the Sco~X-1 data
in the following discussion and other functional forms
could fit the Sco~X-1 data equally well. In
particular, relation~(\ref{scopl}) implies a
non-monotonic change of the peak separation, with a
maximum at an upper kHz QPO frequency of $\sim
700$~Hz. However, we are not proposing here that the
peak separation varies non-monotonically, as this
requires extrapolating relation~(\ref{scopl}) to
frequencies smaller than the lowest kHz QPO frequency
detected so far in Sco~X-1. Figure~3b shows the
observed peak separation of the kHz QPOs in Sco~X-1
together with the curve implied by
relation~(\ref{scopl}). 

The atoll source 4U~1608$-$52 has also shown evidence
for a peak separation that decreases with increasing
kHz QPO frequency (M\'endez et al.\ 1998).  In this
source only one kHz QPO has so far been detected in
the power spectra (see, e.g., Berger et al.\ 1996). 
However, by shifting the power spectra so that the kHz
QPO peaks are aligned and adding them up reveals the
existence of the second, higher-frequency kHz QPO peak
(M\'endez et al.\ 1998). Because of the shifting of
the power spectra and the alignment of the QPO peaks,
the information regarding the centroid frequency of
the upper kHz QPO is lost. Moreover, the ranges of
frequencies of the lower kHz QPOs used originally with
the above technique were overlapping. As a result we
cannot directly compare the decrease of the peak
separation in Sco~X-1 implied by
relation~(\ref{scopl}) to the data of 4U~1608$-$52. 

In the original analysis of M\'endez et al.\ (1998)
the frequency of the lower kHz QPO in the observation
of March 3, 1996 varied in the range $823-893$~Hz. The
peak separation was found to be $232.7\pm 11.5$~Hz. We
have reanalyzed the data of March 6, 1996 using only
the part of the dataset in which the frequency of the
lower kHz QPO was in the range $649-760$~Hz to avoid
any overlap with the range of frequencies detected on
March 3. Even with this restricted part of the dataset
we detected a second QPO peak (at $4.3\sigma$) with a
peak separation equal to 288.1$\pm$11.3~Hz. Figure~3b
compares the 4U~1608$-$52 and Sco~X-1 data; because of
the technique used for 4U~1608$-$52, the March 3 and 6
observations are represented by bands of constant peak
separation. Figure~3b suggests that the decreasing
peak separation found in 4U~1608$-$52 is consistent
with that of Sco~X-1. 

\begin{figure}[t]
 \centerline{
\psfig{file=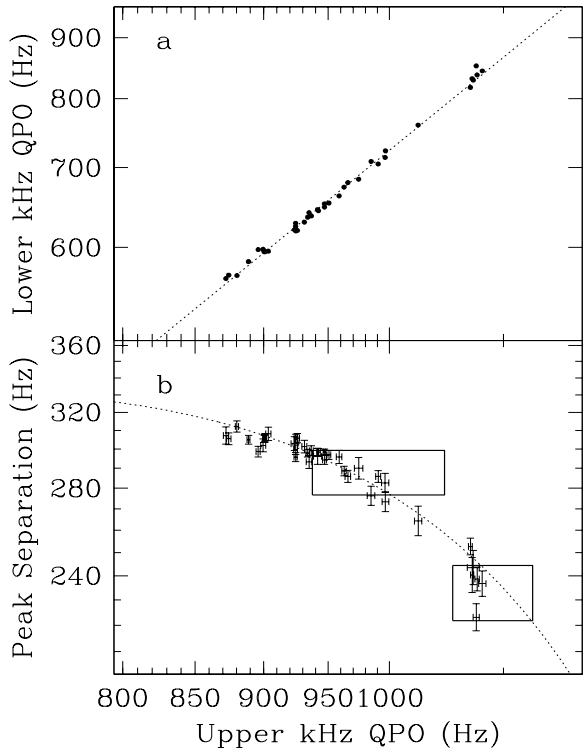,angle=0,height=10truecm,width=8truecm}}
 \figcaption[] {(a) Observed correlation between the
lower and upper kHz QPO frequencies in Sco~X-1. (b)
Observed peak separation between the kHz QPOs in
Sco~X-1 (error bars) and in 4U~1608$-$52 (bands;  see
text). The error bars in panel (a) are smaller than
the size of the dots.}
\end{figure}

The frequencies of the kHz QPOs observed in the
sources used to test the beat-frequency hypothesis in
Figure~2a are {\em also\/} consistent with a
non-constant peak separation, similar to the one
implied by relation~(\ref{scopl}) for Sco~X-1.
Figure~2b demonstrates this by showing the result of
testing the latter hypothesis, i.e., of a power-law
correlation between the lower and upper kHz QPO
frequencies, by means of a $\chi^2$ test. The minimum
$\chi^2$ values for all sources that were found to be
consistent also with a constant peak separation (see
Figure~2a) are $\lesssim 1.0$. 

As mentioned earlier, for all the sources for which
simultaneous kHz QPOs have been detected besides the
ones used to plot Figure~2, the range of observed kHz
QPO frequencies is narrow. We therefore cannot test
directly whether the peak separations between the kHz
QPOs in these sources vary by amounts comparable to
those seen in Sco~X-1 or not. However, Figure~4a shows
that the frequencies of the lower and upper kHz QPOs
observed in all nine sources are remarkably tightly
correlated and follow fairly closely
relation~(\ref{scopl}) that describes the data of
Sco~X-1; Figure~4b shows the instantaneous peak
separations in all nine sources and compares them with
the varying peak separation of Sco~X-1. The tight
correlation between the kHz QPO frequencies shown in
Figure~4a together with the large uncertainties in the
determination of the peak separation for all sources
besides Sco~X-1 showh in Figure~4b strongly suggests
that the data for {\em all\/} sources are consistent
with the varying peak separation of Sco~X-1. Note,
however, that although all sources are consistent with
a peak separation that is varying by amounts similar
to those seen in Sco~X-1, not all these sources are
consistent with a single relation between the lower
and upper kHz QPO frequencies. In fact, the lower and
upper kHz QPO frequencies for all Z sources, with the
possible exception of GX~17$+$2, are consistent with
relation~(\ref{scopl}) that describes the data of
Sco~X-1 (see Fig.\,2b). On the other hand, the data
for the atoll sources 4U~0614$+$09, KS~1731$-$26, and
4U~1636$-$53 are consistent with a single power-law
relation with the same index as relation~(\ref{scopl})
but a lower normalization ($\sim 695$~Hz; the $\chi^2$
values for testing this relation is $\lesssim 1.5$ in
each of the three sources individually). 

\begin{figure}[t]
\centerline{
\psfig{file=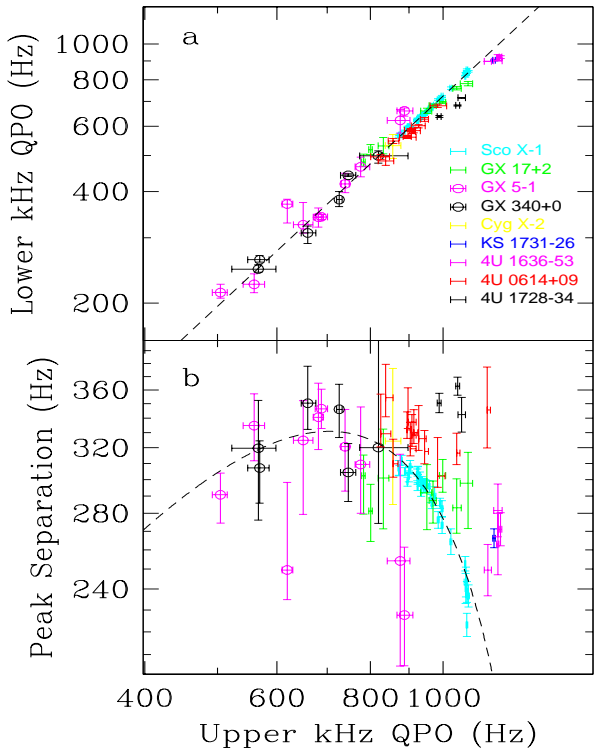,angle=0,height=10truecm,width=8truecm}}
 \figcaption[] {[color plate] (a) Observed correlation 
between the lower and upper kHz QPO frequencies in 
nine LMXBs. (b) Observed peak separation between the 
kHz QPOs in nine LMXBs.}
\end{figure}

\section{DISCUSSION}

In \S2 we showed that, using current kHz QPO data of
neutron-star low-mass X-ray binaries, we cannot reject
the hypothesis that the frequency separations of the
two kHz QPO peaks in the pair are constant in each
source besides Sco~X-1, nor the hypothesis that they
vary in a way similar to Sco~X-1. In Sco~X-1, which is
the only source with very precisely measured centroid
frequencies for the kHz QPOs, the data are
inconsistent with a constant peak separation (see also
van der Klis et al.\ 1997). Furthermore, measurements
of the peak separation in all other sources in which
two simultaneous kHz QPOs have been detected are
consistent with change of the kind observed in
Sco~X-1. This is the case for the Z sources, which are
thought to be accreting at near-Eddington mass
accretion rates (see, e.g., Hasinger \& van der
Klis\markcite{HK89} 1989), as well as for the atoll
sources, which are thought to be accreting at
substantially lower rates. 

These results hint that the frequencies of the upper
and lower kHz QPOs in all sources considered
individually, including Sco~X-1, are consistent with
following a simple (but not necessarily the same)
relation, such as a power-law, which is nevertheless
very similar even for sources with very different mass
accretion rates. Obeying such a relation would
contradict any beat-frequency interpretation of the
pairs of kHz QPOs in LMXBs, in which the frequency
separation of the QPOs is exactly constant. However,
the nearly coherent oscillations observed during
type~I X-ray bursts in two sources with frequencies
closely equal to the peak separations of the kHz QPOs
and the fact that the oscillation amplitudes evolve
systematically during the bursts are very strong
evidence that the peak separations in these sources
are similar to the spin frequencies of the neutron
stars (Strohmayer et al.\ 1996, 1997b). Most
importantly, in 4U~1728$-$34 the frequencies of the
oscillations in the tails of bursts separated by about
20 months are consistent with being constant, implying
a timescale for the frequency change of $\gtrsim
10^3-10^4$~yr (Strohmayer\markcite{S97} 1997). The
only conceivable frequency in these systems that is
stable to the degree inferred from the observations of
4U~1728$-$34 is the spin frequency of the neutron
star. Therefore, the frequency separation of the two
kHz QPO peaks in this source appears to be {\em
closely equal but perhaps not identical\/} to the spin
frequency of the neutron star. In 4U~1636$-$536 and in
KS~1731$-$26, the peak separation of the kHz QPOs also
appears to be directly related to the neutron star
spin frequency. 

In current beat-frequency models for the pair of kHz
QPOs (see, e.g., Strohmayer et al.\ 1996; Miller et
al.\ 1998), the upper kHz QPO peak is produced at the
Keplerian orbital frequency at a characteristic radius
in the accretion disk; the lower kHz QPO peak is then
produced at the beat frequency of the upper kHz QPO
with the neutron star spin. In order for a
beat-frequency model to account for the varying peak
separation between the kHz QPOs, one of the above
assumptions would need to be relaxed. For example, the
frequency that is beating with the neutron star spin
to produce the lower kHz QPO may not be the frequency
of the upper kHz QPO, i.e., the two frequencies could
correspond to different radii in (or heights above)
the disk plane (see Miller et al.\ 1998).  The fact
that the variation in the peak separation of the two
kHz QPOs in Sco~X-1 is larger than the FWHM of either
the lower or upper kHz QPOs implies that the two
annuli or regions in the accretion disk responsible
for the two kHz QPOs are not overlapping.
Alternatively, the frequency that is beating with the
upper kHz QPO to produce the lower kHz QPO may be
nearly but not strictly equal to the neutron star spin
frequency (see, e.g., White \& Zhang\markcite{WZ97}
1997). 

In conclusion, the data from both the Z and atoll
sources are consistent with a varying peak separation
between the kHz QPOs. If future data support this
conjecture, they will pose interesting new constraints
on beat-frequency models for these QPOs: the peak
separation should correlate more strongly with the
frequency of the upper kHz QPO than with the mass
accretion rate or the magnetic field strength. 

\acknowledgements

DP thanks Vicky Kalogera and Deepto Chakrabarty for
many useful discussions and for carefully reading the
manuscript. This work was supported in part by a
post-doctoral fellowship of the Smithsonian Institute
(DP), by a fellowship of the Consejo Nacional de
Investigaciones Cientificas y T\'ecnicas de la
Rep\'ublica Argentina (MM), by the Netherlands
Organization for Scientific Research (PWO) grant PCS
78-277 and the Netherlands Foundation for Research in
Astronomy (ASTRON) grant 781-76-017 (RW, JH, PJ,
MvdK), by NSF grant AST~96-18524 (FKL), by NASA grants
NAG~5-2925 (FKL), NAG~5-2868 (MCM), NAG~5-3269 and
NAG~5-3271 (JvP), and by several {\it RXTE\/}
observing grants. WHGL also gratefully acknowledges
support from NASA. 


\end{document}